\documentclass[conference,letterpaper]{IEEEtran}

\usepackage[utf8]{inputenc} 
\usepackage[T1]{fontenc}
\usepackage{url}
\usepackage{ifthen}
\usepackage{cite}
\usepackage[cmex10]{amsmath} 

\DeclareMathOperator*{\argmin}{\arg\!\min}
\DeclareMathOperator{\vol}{vol}

\usepackage{amssymb}
\usepackage{nicefrac}
\usepackage{xcolor}
\usepackage{graphicx}

\newtheorem{theorem}{Theorem}
\newtheorem{corollary}{Corollary}[theorem]
\newtheorem{lemma}[theorem]{Lemma}
\newtheorem{proposition}[theorem]{Proposition}
\newtheorem{definition}{Definition}

\newcommand{\hatx}{\ensuremath{\hat{x}}}
\newcommand{\hatX}{\ensuremath{\hat{X}}}
\DeclareMathOperator{\supp}{supp}
\DeclareMathOperator{\sgn}{sgn}

\renewcommand{\baselinestretch}{0.964}

\begin{document}
\title{Critical Slowing Down Near Topological Transitions in Rate-Distortion Problems} 

\IEEEoverridecommandlockouts
\author{%
  \IEEEauthorblockN{Shlomi Agmon\IEEEauthorrefmark{1}\IEEEauthorrefmark{2},
                    Etam Benger\IEEEauthorrefmark{1}\IEEEauthorrefmark{2},
                    Or Ordentlich\IEEEauthorrefmark{2},
                    and Naftali Tishby\IEEEauthorrefmark{2}\IEEEauthorrefmark{3}}
  \IEEEauthorblockA{\IEEEauthorrefmark{2}%
                    School of Computer Science and Engineering,
                    The Hebrew University of Jerusalem,
                    Jerusalem,
                    Israel}
  \IEEEauthorblockA{\IEEEauthorrefmark{3}%
                    Edmond and Lily Safra Center for Brain Sciences,
                    The Hebrew University of Jerusalem,
                    Jerusalem,
                    Israel}
  \IEEEauthorblockA{Email: \{shlomi.agmon, etam.benger, or.ordentlich, naftali.tishby\}@mail.huji.ac.il}
  \thanks{\IEEEauthorrefmark{1} These authors contributed equally to this work.}
  \thanks{This work was supported in part by the Gatsby Foundation and in part by the Pazy Foundation. E.B.\ is further supported by the Center for Interdisciplinary Data Science Research (CIDR) at the Hebrew University. The work of O.O.\ was supported by the ISF under Grant 1791/17.}
}

\maketitle

\begin{abstract}
In rate-distortion (RD) problems one seeks reduced representations of a source that meet a target distortion constraint. Such optimal representations undergo topological transitions at some critical rate values, when their cardinality or dimensionality change. We study the convergence time of the Arimoto-Blahut alternating projection algorithms, used to solve such problems, near those critical points, both for the rate-distortion and information bottleneck settings. We argue that they suffer from critical slowing down -- a diverging number of iterations for convergence -- near the critical points. This phenomenon can have theoretical and practical implications for both machine learning and data compression problems.    
\end{abstract}


\section{Introduction}

Given a source $X \sim p(x)$ on a finite alphabet $\mathcal X$, a representation alphabet $\mathcal \hatX$, and a distortion measure $d: \mathcal X \times \mathcal \hatX \to \mathbb R^+$, the rate-distortion function (RDF) is defined as $R(D) = \min I(X; \hatX)$, where the minimization is with respect to all test channels $p(\hatx | x)$ satisfying the distortion constraint $\mathbb E [d(X, \hatX)] \leq D$,~\cite{berger71,Cover2006}.
The distortion-rate function $D(R)$ is merely the inverse of $R(D)$. An analytic expression for $R(D)$ (or $D(R)$) involves solving the minimization above, and is only known for some special cases. However, it is possible to obtain a numerical solution using different algorithms, including the Arimoto-Blahut (AB) algorithm~\cite{arimoto1972,blahut1972}.

Clearly, for $R=0$, the test channel that maps all $x \in \mathcal X$ to $\argmin_{\hatx} \mathbb E [d(X, \hatx)]$ is optimal, whereas for $R=H(X)$ the channel that maps any $x \in \mathcal X$ to $\argmin_{\hatx} d(x,\hatx)$ is optimal. Thus, the cardinality (support size) of the optimal $\hatX$ attaining $D(R)$ changes as we increase/decrease $R$. Typically, the cardinality of the optimal $\hatX$ decreases gradually from $|\mathcal \hatX| = |\mathcal X|$ to $|\mathcal \hatX| = 1$ as we decrease $R$ from $H(X)$ to $0$, but there are also examples where the cardinality of the optimal $\hatX$ behaves non-monotonically~\cite[Section 2.7]{berger71}. We refer to these changes of the representation cardinality as \emph{topological-} or \emph{phase transitions} and the values of $R$ where they occur as \emph{critical}.
This paper studies the algorithmic difficulty of computing $D(R)$ near such critical values of $R$.

In the context of data compression, the main quantity of interest is the representors' distribution, $p(\hat{x})$, at a given value of $R$, for which an optimal code is constructed and the convergence to the optimal $D(R)$ can be obtained, as the blocklength increases. In applications of RD to machine learning and statistical physics, however, there is more interest in the nature of the optimal channel, or \emph{representation encoder}, $p(\hat{x}|x)$. The reason is that in machine learning the similar features of the source patterns $x$ that are mapped to specific representations $\hat{x}$ determine the relevant order parameters and topology of the problem. 
At the critical points, the neighborhoods of patterns can merge or split during the learning process, and the induced topology of the patterns -- which patterns are neighbours  -- can significantly alter. Understanding such topological changes and the nature of the encoder is critical in representation learning~\cite{bengio2014representation} and has gained recent interest also in statistical physics~\cite{gordon2020relevance}.

Similar topological transitions occur also in the closely related approach of the information bottleneck (IB)~\cite{IB1999}, which aims to achieve maximal compression of $X$ while preserving most \emph{relevant information} about another correlated variable $Y$. This approach has recently drawn attention due to its possible relation to the learning dynamics of deep neural networks (e.g.~\cite{tishby2015deep,shwartzziv2017opening, Bachrach2003}). Specifically, there is evidence to suggest that the representations of the layers in such networks converge to successively refineable points near the IB curve, which may be related to the critical points where such transitions occur~\cite{shwartzziv2017opening,goldfeld2020information}.
Moreover, the critical points represent changes in the nature of the optimal solutions to the RD or IB problems, when considering the size of the representation alphabet as an additional constraint. 

In this work we show that solutions to RD problems lose their stability 
at the critical points. As a result, we prove that the AB algorithm 
slows down dramatically near such critical points. This phenomenon, in which systems' dynamics slow down near phase transitions, is known in statistical physics as \emph{critical slowing down} (CSD). Finally, we show that similar slowing down occurs also for the extended AB algorithm, used to solve IB problems numerically, near critical points. 

\begin{figure*}[htbp]
  \centering
  \includegraphics[width=0.9\textwidth]{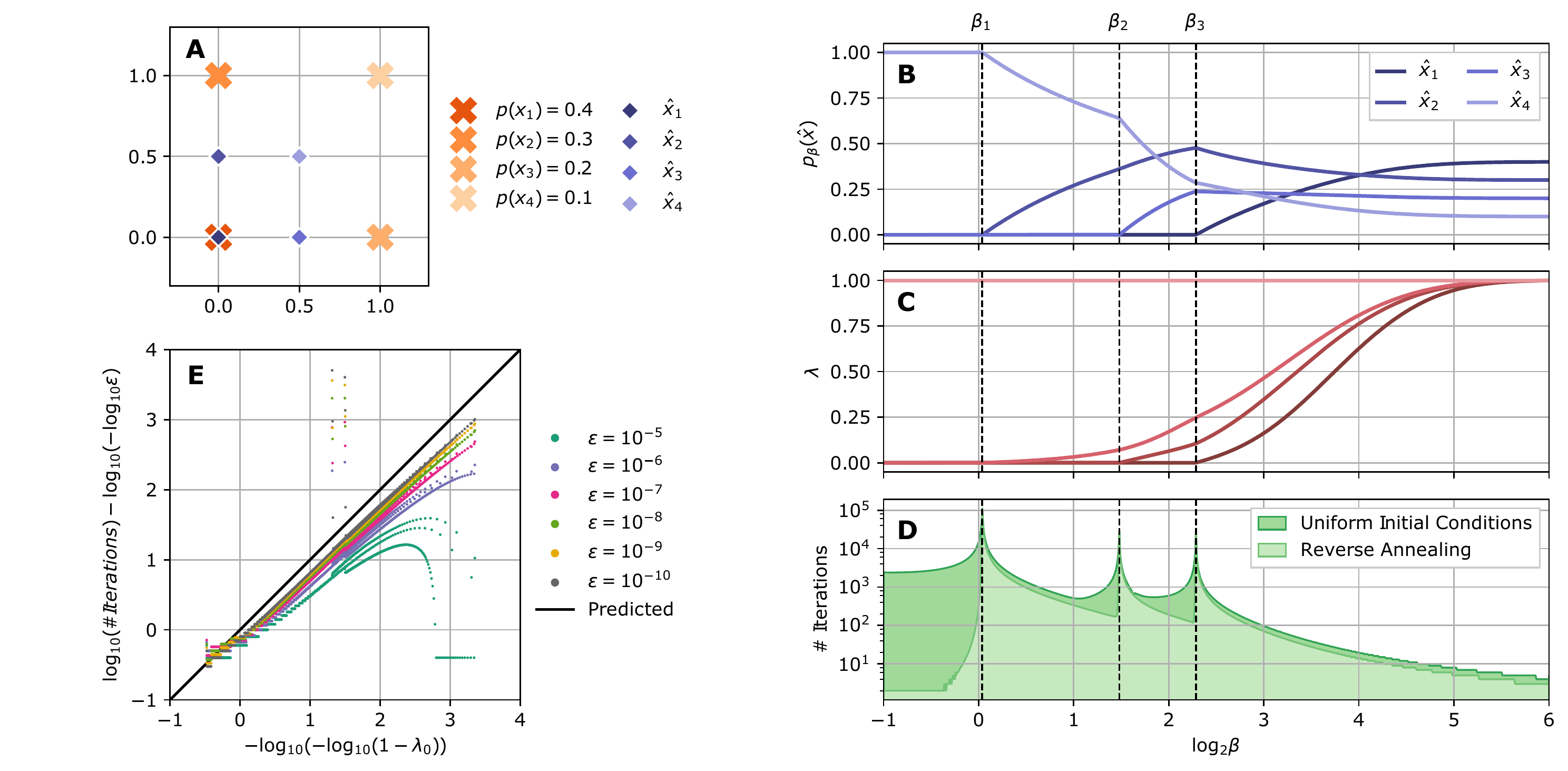}
  \caption{\textbf{A.} A simple RD problem: the sets $\mathcal X, \mathcal \hatX \in \mathbb R^2$ shown in orange and purple, correspondingly; $p(x) = (0.4, 0.3, 0.2, 0.1)^\intercal$; the distortion function is defined as $d(x, \hatx) = \frac{1}{\mu} || x - \hatx ||_2^2$, where $\mu = \max_{x' \in \mathcal X,\ \hatx' \in \mathcal \hatX} || x' - \hatx' ||_2^2$ is a normalization factor.
  \textbf{B.} Values of $p_\beta(\hatx)$, the solutions to the RD problem at different values of $\beta$: note the three phase transitions at $\beta_1,\,\beta_2,\,\beta_3$ (marked with dashed lines) -- until $\beta_1$ only $p(\hatx_4) > 0$, then at $\beta_1$ the representor $\hatx_2$ starts to gain mass, and at $\beta_2$ also $\hatx_3$, finally $\hatx_1$ starts to gain mass at $\beta_3$. \textbf{C.} Eigenvalues of $A$ at $p_\beta$: note that near each of the phase transitions an eigenvalue of $A$ approaches 0. \textbf{D.} Number of iterations until convergence ($\varepsilon = 10^{-9}$, logarithmic scale) using uniform initial conditions and reverse annealing: a slowing down of approximately an order of magnitude is clearly noticed at each critical point.
  \textbf{E.} The relation between the number of iterations until convergence and $\lambda_0$, the smallest nonzero eigenvalue of $A$, computed using reverse annealing with various values of $\varepsilon$: as $\varepsilon$ decreases, the relation approaches the limit formula in Theorem~\ref{thm:worst-case-convergence-time-for-AB} (diagonal line); the artifact at the center, consisting of a few vertically arranged points, corresponds to the slowing down to the left of $\beta_1$ (we do not fully understand this phenomenon, which is more prominent in the uniform initial conditions setting).}
  \label{fig:slowingdown}
\end{figure*}

\section{Related Work}
The convergence of the AB algorithm for finite/countable reconstruction alphabets was established in~\cite{blahut1972,arimoto1972,Csiszar74}. Boukris~\cite{boukris1973upper} further derived an upper bound on the convergence rate, which shows that the gap between the value of the Lagrangian defined below in~\eqref{eq:Largange} under the AB output and the optimal solution decreases at least inversely proportional to the number of iterations. Several papers have analyzed the convergence rate of the AB algorithm for capacity computation, and have demonstrated that the algorithm converges exponentially fast whenever the support of a capacity achieving input distribution is full~\cite{Yu10,nakagawa2020analysis,md04}.

Phase transitions in the optimal test-channel attaining $D(R)$, as $R$ changes, were already discussed by Berger~\cite 
{berger71}. In fact, Berger also showed that if the support of the optimal $\hat{X}$ is known, the computation of the optimal test-channel simplifies. In a sense, this already shows that finding an optimal test-channel at critical points, where the support of the optimal solution changes, can be computationally challenging.
Two decades later, Rose~\cite{rose94} demonstrated that even when the reconstruction alphabet is continuous, the optimal $\hat{X}$ typically has finite support, which grows with $R$.

The extended version of the AB algorithm for the IB problem (henceforth, the IB algorithm), in which the representors are optimized as well, was proven in~\cite{IB1999} to converge, although not necessarily to a unique minimum as the convexity is lost. In~\cite{SlonimAgglomerative2000}, the authors address the difficulty to identify the topological transitions in the IB framework using the method of deterministic annealing~\cite{rose1990deterministic}. To solve this difficulty, Parker et al.~\cite{ParkerAnnealing2003,GedeonMathematical2012} study the bifurcation structure of solutions to the IB and other RD-like problems using bifurcation theory. 
They focus mainly on the first critical point -- where they argue that the trivial solution (i.e.\ $|\mathcal \hatX| = 1$) loses its stability and structure begins to emerge. More recently, phase transitions in the IB have been studied from a representation learning perspective, shown to relate to learning of new features in the data~\cite{WuLernability2020,WuPhase2020}, and algorithms for finding the critical points were presented~\cite{ZaslavskyTishby2019}. Analytical expression for the location of the critical points is known in the Gaussian IB case~\cite{ChechikGIB2005, goldfeld2020information}, where the topological transitions correspond to changes in the dimension of the Gaussian distribution of $\hatX$.

\section{Critical Points in Rate-Distortion Theory}
The constrained optimization problem of RD is solved by introducing a positive Lagrange multiplier, $\beta$, to impose the expected distortion constraint, and minimizing the Lagrangian~\cite{Cover2006}
\begin{align}
  I(X; \hatX) + \beta\,\mathbb E [d(X, \hatX)],
  \label{eq:Largange}
\end{align}
with respect to all test channels $p(\hatx | x)$.
The Lagrange multiplier $\beta$, of a role similar to that of inverse temperature in statistical physics, 
determines the topological structure of the optimizing channel as well as the trade-off between compression and distortion. 
A solution to the RD problem at a given value of $\beta$, denoted by $p_\beta(\hatx | x)$, that is, a minimizer of~\eqref{eq:Largange},
%
%
must (self-consistently) satisfy both equations~\cite{arimoto1972,
berger71,
Cover2006}:
\begin{align}
    p_\beta(\hatx | x) &= \frac{p_\beta(\hatx) e^{-\beta d(x, \hatx)}}{Z(x, \beta)} \label{eq:p(hatx|x)}\\
	\text{and} \quad p_\beta(\hatx) &= \sum_x p_\beta(\hatx | x) p(x) \;, \label{eq:p(hatx)}
\end{align}
where $Z(x, \beta) := \sum_{\hatx} p_\beta(\hatx) e^{-\beta d(x, \hatx)}$ is a normalization (partition) function. 
Since $p_\beta(\hatx |x)$ is determined by $ p_\beta(\hatx)$ when $\beta$ and the distortion function are given, 
we may consider $p_\beta(\hatx)$ as the optimization variable. Moreover, to simplify the discussion, we assume throughout that $p_\beta(\hatx)$ is unique for all values of $\beta$, and consider it as a vector $p_\beta$ in the sequel.
\begin{definition} \label{def:phasetransition}
  Denote the support of a solution $p_\beta$ by $\supp p_\beta = \{\hatx \in \mathcal \hatX : p_\beta(\hatx) > 0\}$.
  We say that $\beta_c$ is \emph{critical} and that the RD problem has a \emph{topological transition} at $\beta_c$ if
  \begin{equation}
    | \supp p_{\beta^-} | \neq | \supp p_{\beta^+} |
  \end{equation}
  for all $\beta^- < \beta_c < \beta^+$ in some (small) neighborhood of $\beta_c$. 
\end{definition}

We restrict our discussion to local stability analysis, and thus assume that $\beta \mapsto p_\beta$ is continuous. As a result, we tackle only what is known in statistical physics as phase transitions of second order or higher.

Equations~\eqref{eq:p(hatx|x)} and~\eqref{eq:p(hatx)} above can be written concisely as $F=0$, where for all $\hatx \in \mathcal \hatX$
\begin{equation}
	\big[F( p, \beta )\big]_{\hatx} := p(\hatx) - \sum_x p(x) \frac{p(\hatx) e^{-\beta d(x, \hatx)}}{Z(x, \beta)} \;. \label{eq:F}
\end{equation}
In what follows, we study the properties of $F$ and its Jacobian $\nabla F = \nicefrac{\partial \left[F\left( p, \beta \right)\right]_{\hatx}}{\partial p(\hatx')}$ 
around critical points, and provide a characterization of the phase transitions of RD problems in terms of $\nabla F$. This characterization will play a pivotal role, as we show next that $\nabla F$ is closely related to the Jacobian of a single step of the Arimoto-Blahut algorithm and its spectrum governs the algorithm's convergence rate.


Assume there is a topological transition at $\beta_c$, such that $|\supp p_{\beta^-}| < |\supp p_{\beta^+}|$ in the above notation, then there exists $\hatx_0$ such that $p_{\beta^-}(\hatx_0) = 0$ and $p_{\beta^+}(\hatx_0) > 0$.
Consider the representation space $\mathcal \hatX' = \mathcal \hatX \setminus \{\hatx_0\}$ and the distortion function $d'$, which is the restriction of $d$ to $\mathcal \hatX'$. 
Clearly, the solution $q_\beta$ of the RD problem defined by $d'$ is identical to $p_\beta$ (restricted to $\mathcal \hatX'$) at a left neighborhood of $\beta_c$, but they must differ to the right. Nevertheless, the extension of $q_\beta$ over the original problem satisfies $F(q_\beta, \beta)=0$ also to the right of $\beta_c$, if one sets $q_\beta(\hatx_0) = 0$. While this would not be a solution to the RD problem, it shows the number of solutions to $F=0$ changes at critical values of $\beta$, or that $F=0$ has a \emph{bifurcation} at $\beta_c$.

Although the extension of $q_\beta$ is a fixed point of the AB algorithm, it is not a stable solution, as adding a small perturbation to $q_\beta(\hatx_0)$ leads to convergence to the optimal solution $p_\beta$ instead. Hence, at $\beta_c$ the existing solution to the RD problem loses its stability with respect to the AB algorithm, and a new stable solution emerges. 

Notice that $\nabla F$ must be singular at critical values of $\beta$.
Otherwise, by the implicit function theorem, there must exist a unique solution to $F=0$ as a function of $\beta$ in the vicinity of $\beta_c$~\cite{kielhofer2012}. Unfortunately, $\nabla F$ is trivially singular when $p_\beta(\hatx) = 0$ for some representor $\hat{x}$, so this is not a useful characterization of the critical points. However, we show next that the Jacobian becomes "more singular" with each transition.

Let $A$ denote the transposed Jacobian matrix of $F$ at a solution $p_\beta$, then~(see Appendix~\ref{sec:A})
\begin{align}
	A_{\hatx \hatx'} &:= \big[(\nabla F)^\intercal\big]_{\hatx \hatx'} = \sum_x p_\beta(\hatx' | x) p_\beta(x | \hatx) \label{eq:A} \\
	&= p_\beta(\hat x') \sum_x p(x) \frac{e^{-\beta \big( d(x, \hatx) + d(x, \hatx') \big)}}{Z(x, \beta)^2} \;, \label{eq:A2}
\end{align}
and we have the following result:
\begin{theorem} \label{thm:dimker}
	Let $m = |\mathcal \hatX|$ and $k = |\supp p_\beta|$, then at $p_\beta$ $\dim \ker A = m - k$.
\end{theorem}

\begin{corollary}
	Topological transitions of a RD problem occur exactly at values of $\beta$ where the dimension of $\ker A$, at the solutions $p_\beta$, changes.
\end{corollary}

The proof
consists of two steps~(see Appendix~\ref{sec:proofThmDimKer}). First, we show that $\dim \ker A \geq m - k$:
\begin{lemma} \label{lemma:standardbasis}
	Let $\mathbf e_{\hat x} \in \mathbb R^m$ denote the standard basis vector with 1 at the $\hatx$ coordinate and 0 elsewhere, then $p_\beta(\hat x) = 0 \iff \mathbf e_{\hat x} \in \ker A$.
\end{lemma}

Second, assume for simplicity that $d$ is finite and without loss of generality $d(\cdot, \hatx_1) \neq d(\cdot, \hatx_2)$ for all $\hatx_1 \neq \hatx_2$. Then there is always a standard basis of $\ker A$, resulting in the converse inequality $\dim \ker A \leq m - k$:
\begin{proposition} \label{prop:allbasis}
	Let $v \in \ker A$, such that exactly $r \geq 1$ of its coordinates, denoted $\hatx_1, \dots, \hatx_r$, are nonzero; then all the corresponding standard basis vectors $\mathbf e_{\hatx_i}$, for $1 \leq i \leq r$, belong to $\ker A$.
\end{proposition}

Theorem~\ref{thm:dimker} refers to the geometric multiplicity of the eigenvalue 0 of $A$, and shows that it changes at critical points. However, to ensure that near such points $A$ must have a small positive eigenvalue we need to establish a similar statement for the algebraic multiplicity. This will follow as a corollary of the next theorem~(see Appendix~\ref{sec:diagonalizable}):
\begin{theorem} \label{thm:diagonalizable}
    The matrix $A$ is diagonalizable with real non-negative eigenvalues.
\end{theorem}

Together with Theorem~\ref{thm:dimker} above, we obtain:
\begin{corollary} \label{cor:algebraic}
    Let $m = |\mathcal \hatX|$ and $k = |\supp p_\beta|$, then at $p_\beta$ the \emph{algebraic} multiplicity of the eigenvalue 0 of $A$ is exactly $m - k$.
    Consequently, topological transitions of a RD problem occur exactly at values of $\beta$ where the algebraic multiplicity of the eigenvalue $0$ of $A$ changes.
\end{corollary}

Figure~\ref{fig:slowingdown} demonstrates this result on a simple RD problem, consisting of 4 points in $\mathbb R^2$ (Figure~\ref{fig:slowingdown}A). The solutions to the problem undergo 3 transitions, at $\beta_1, \beta_2$ and $\beta_3$, where the cardinality of $\hatX$ increases from 1 (trivial solution) to 2, 3 and 4, correspondingly (Figure~\ref{fig:slowingdown}B). Figure~\ref{fig:slowingdown}C shows that at each critical $\beta$ another eigenvalue of $A$ reaches 0.

Finally, as the matrix $A$ is row-stochastic, all its eigenvalues are inside the unit circle~\cite{horn2012matrix}, and by Theorem~\ref{thm:diagonalizable}, they are in $[0, 1]$, as required in the next section.

\section{Slowing Down of Arimoto-Blahut}
\label{sec:slowing-down-of-AB}

The numerical computation of solutions to RD problems is usually performed using the Arimoto-Blahut (AB) algorithm~\cite{arimoto1972,blahut1972}. It consists of an alternating minimization, applying Equations~\eqref{eq:p(hatx|x)} and \eqref{eq:p(hatx)} repeatedly, starting from some initial distribution $p_0$ in the interior of the simplex $\Delta \mathcal \hatX$. The $k$-th iteration $p_k$ of the AB algorithm is said to have \textit{$\varepsilon$-converged} to a RD solution $p_\beta$ if
\footnote{ The exact norm used at \eqref{eq:eps-convergence-def} is of little importance, as $|\mathcal{\hat{X}}|$ is finite and we are typically interested in small values of $\varepsilon$. For convenience, we have chosen $L^1$ in Theorem~\ref{thm:worst-case-convergence-time-for-AB} below, and $L^\infty$ in the Figures~\ref{fig:slowingdown} and~\ref{fig:slowingdownIB}.}
\begin{equation} \label{eq:eps-convergence-def}
	\| p_k - p_\beta \| < \varepsilon \;.
\end{equation}

Given a value of $\beta$, define the operator $AB: \Delta \mathcal \hatX \rightarrow \Delta \mathcal \hatX$ to be the result of applying a single step of the $AB$ algorithm. Solutions to the RD problem are fixed points of the algorithm, that is $AB\, p_\beta = p_\beta$ or $\big( I - AB \big) p_\beta = 0$, where $I$ is the identity operator. Using \eqref{eq:F} it follows that at the solutions of the RD problem $AB = I - F$, hence $\nabla AB \rvert_{p_\beta} = I - A^\intercal$, where $\nabla AB\rvert_{p_\beta}$ is the Jacobian matrix of $AB$ at $p_\beta$.

Let $\delta p_k = p_k - p_\beta$ be the deviation from $p_\beta$, then to first order in $\delta p_k$, 
\begin{gather}
	AB\, p_k \approx p_\beta + \nabla AB \rvert_{p_\beta} \delta p_k\label{eq:linearization1} \\ \nonumber
	\Rightarrow\; \delta p_{k + 1} = AB\, p_k - p_\beta \approx \big( I - A^\intercal \big) \delta p_k \;.
\end{gather}
Hence,
\begin{equation} \label{eq:first-order-approx-of-repeated-AB}
	\delta p_{k} \approx \big( I - A^\intercal \big)^k \delta p_0 \;.
\end{equation}
As a result, the convergence rate of the algorithm is governed by the largest eigenvalue of $I - A^\intercal$ inside the unit circle, denoted $\lambda_{max}$. If the initial deviation $\delta p_{0}$ has a nonzero component in the eigenspace of $\lambda_{max}$, then
\begin{equation}		\label{eq:linear-bound-on-iterations}
	\| \delta p_{k} \| < \varepsilon \quad \overset{\eqref{eq:first-order-approx-of-repeated-AB}}{\Longrightarrow} \quad
	k \approx \frac{-\log \varepsilon + \text{const}}{-\log |\lambda_{max}|} \;.
\end{equation}
For the asymptotic convergence rate we consider $\lim_{\varepsilon\to 0} \frac{k}{-\log \varepsilon}$, to avoid dependence on the particular choice of initial conditions, via the constant at \eqref{eq:linear-bound-on-iterations}.


This argument is made precise by the next theorem, proven in Appendix~\ref{sec:worst-case-convergence-time-for-AB}. 
Recall that $A$ is diagonalizable (Theorem~\ref{thm:diagonalizable}) and its eigenvalues are in $[0, 1]$. 
Therefore, $\nabla AB = I - A^\intercal$ is also diagonalizable and its eigenvalues are in $[0, 1]$. When $p_\beta$ has a full support, the eigenvalues of $\nabla AB$ are in $[0, 1)$ (Theorem~\ref{thm:dimker}). Consequently, we have $\lambda_{max} = 1 - \lambda_0 < 1$, where $\lambda_0 > 0$ is the smallest nonzero eigenvalue of $A$.
\begin{theorem} \label{thm:worst-case-convergence-time-for-AB}
	Let $p_\beta$ be a RD solution with $p_\beta(\hatx) > 0$ for all $\hatx$, and $\lambda_{max} = 1 - \lambda_0 < 1$ the largest eigenvalue smaller than 1 of $\nabla AB$ at $p_\beta$. Denote by $k(p_0, \varepsilon)$ the number of iterations required for an initial distribution $p_0$ to $\varepsilon$-converge to $p_\beta$, and define $B(\delta) := \{ p \in \Delta\mathcal\hatX : \|p - p_\beta\|_1 \leq \delta \}$. Then, for any $a > 0$,
	\begin{equation}		\label{eq:asymptotic-bound-on-convergence-rate-in-thm}
		\Pr_{p_0 \sim \mathcal{U}\left(B(\delta)\right)} \left( 
		\left|\lim_{\varepsilon \to 0^+} \frac{k(p_0, \varepsilon)}{-\log \varepsilon} - \frac{1}{-\log 		\lambda_{max}}\right| < a
		\right) \underset{\delta\to 0}{\longrightarrow} 1 \;,
	\end{equation}
	where $\mathcal{U}(S)$ denotes the uniform distribution on $S$.
\end{theorem}

A lower bound cannot be expected to hold for \textit{every} initial condition in the vicinity of $p_\beta$. Indeed, even if the linearization in~\eqref{eq:linearization1} were exact, the lower bound~\eqref{eq:linear-bound-on-iterations} holds for all but a zero-measure set of initial conditions -- those with no component in the $\lambda_{max}$ eigenspace. In the general case, where the dynamics are nonlinear, the fraction of applicable initial conditions increases in the vicinity of $p_\beta$ as $\delta \to 0$, \eqref{eq:asymptotic-bound-on-convergence-rate-in-thm}. The rate at which this fraction approaches 1 depends on the desired accuracy $a$. 

Consider a topological transition of the RD problem at $\beta_c$, such that $|\supp p_{\beta^-}| < |\supp p_{\beta^+}|$. 
According to Corollary~\ref{cor:algebraic}, the algebraic multiplicity of the eigenvalue $0$ of $A$ is greater at $\beta^-$ than at $\beta^+$. Since $A$ is continuous in $p_\beta$ and $p_\beta$ is assumed continuous in $\beta$, 
there exists an eigenvalue $\lambda_0 > 0$ of $A$ such that $\lambda_0 \to 0$ as $\beta$ approaches $\beta_c$ from above. When coordinates $\hat{x}$ outside $\supp p_{\beta^+}$ are initialized at 0, as in reverse annealing (see below), AB effectively coincides with its restriction to $\supp p_{\beta^+}$. 
Consequently, by Theorem~\ref{thm:worst-case-convergence-time-for-AB}, the AB algorithm experiences a significant slowing down 
as it gets closer to the critical point.

This is clearly observed in simulations (Figure~\ref{fig:slowingdown}D) where the number of iterations until convergence is plotted in two different settings: reverse annealing and uniform initial conditions,\footnote{Other similar choices in the interior of $\Delta \mathcal \hatX$, e.g.\ sampling according to the symmetric Dirichlet distribution $p_0 \sim Dir(\mathbf 1)$, do not yield significantly different results.} $p_0(\hatx) = \nicefrac{1}{| \mathcal \hatX |}$. In reverse annealing the algorithm is run for decreasing values of $\beta$, starting from the converged solution at the previous $\beta$. 
In both settings there is a noticeable slowing down to the right of the critical points, as expected from Theorem~\ref{thm:worst-case-convergence-time-for-AB}, given the small nonzero eigenvalues of $A$ in those areas (Figure~\ref{fig:slowingdown}C).

In addition, it can be seen that the reverse annealing method always converges faster than uniform initial distributions. 
In the latter, there is an overall increase in the iterations baseline as $\beta$ decreases,
since more eigenvalues reach 0.

Finally, Figure~\ref{fig:slowingdown}E shows that the number of iterations required to converge to a solution within accuracy $\varepsilon$ gets closer to the bound in Theorem~\ref{thm:worst-case-convergence-time-for-AB} as $\varepsilon$ decreases. Smaller $\varepsilon$ are needed when approaching a topological transition, as can be seen by examining the second order terms (the details are omitted).

\section{Critical Slowing Down in the IB Framework}
Given a pair of random variables $(X,Y) \sim p(x,y)$, such that $I(X; Y) > 0$, the information bottleneck approach aims to find a channel $p(\hatx | x)$ that minimizes $I(X; \hatX)$, while preserving as much information $I(\hatX;Y)$ as possible~\cite{IB1999}.
While the IB problem can be viewed as a noisy source coding problem where the reconstruction alphabet is $\Delta\mathcal\hatX$, the AB algorithm does not directly apply, since $\mathcal\hatX$ is continuous. Nevertheless, it is known~\cite{harremoes2007information} that for each $\beta$, taking at most $|\mathcal{X}|$ points of the simplex $\Delta\mathcal\hatX$ suffices. If one were given those points of the simplex, the problem would be reduced to a standard RD problem on a finite reconstruction alphabet. However, since those are not known \emph{a priori}, the IB problem can be thought of as an envelope of many different RD problems, making the optimization problem non-convex\cite{Bachrach2003}.


As in RD, the constrained optimization problem of the IB is solved by minimizing a Lagrangian, 
\begin{equation}
    I(X; \hatX) - \beta\,I(\hatX; Y)
\end{equation}
over all channels from $\mathcal{X}$ to $\mathcal{\hatX}=\{1,\ldots,|\mathcal{X}|\}$.
Consequently, solutions to the IB problem follow a set of self consistent equations, similar to those of RD (Equations~\eqref{eq:p(hatx)} and \eqref{eq:p(hatx|x)}) with the addition of the \emph{decoder} equation
\begin{equation}
	p_\beta(y | \hatx) = \sum_x p(y | x) \frac{p_\beta(\hatx | x) p(x)}{p_\beta(\hatx)} \;, \label{eq:decoder}
\end{equation}
and the distortion function in \eqref{eq:p(hatx|x)} is given by $d_{IB}(x,\hatx)=D_{K\!L}[p(y|x) || p_\beta(y|\hatx)]$. Notice that this distortion depends (indirectly) on the encoder distribution and on $\beta$. Moreover, here $p_\beta(\hatx)$ does not capture the entire solution, as in RD, which must be described by $p_\beta(\hatx | x)$. As a result, the analysis of the IB topological transitions is somewhat more complicated. 

In addition, the transitions in the IB framework do not consist only of changes in the size of the support, $\supp p_\beta$. In fact, various representors may share the same decoder distribution $p_\beta(y | \hatx)$, making them essentially equivalent. The relevant quantity that changes in topological transitions of the IB is the \emph{effective cardinality}~\cite{ZaslavskyTishby2019}, defined as the number of \emph{different} non-empty decoder distributions, $|\{p_\beta(y | \hatx) : p_\beta(\hatx) > 0\}|$.

\begin{figure}[htbp]
	\centering
	\includegraphics[width=0.45\textwidth]{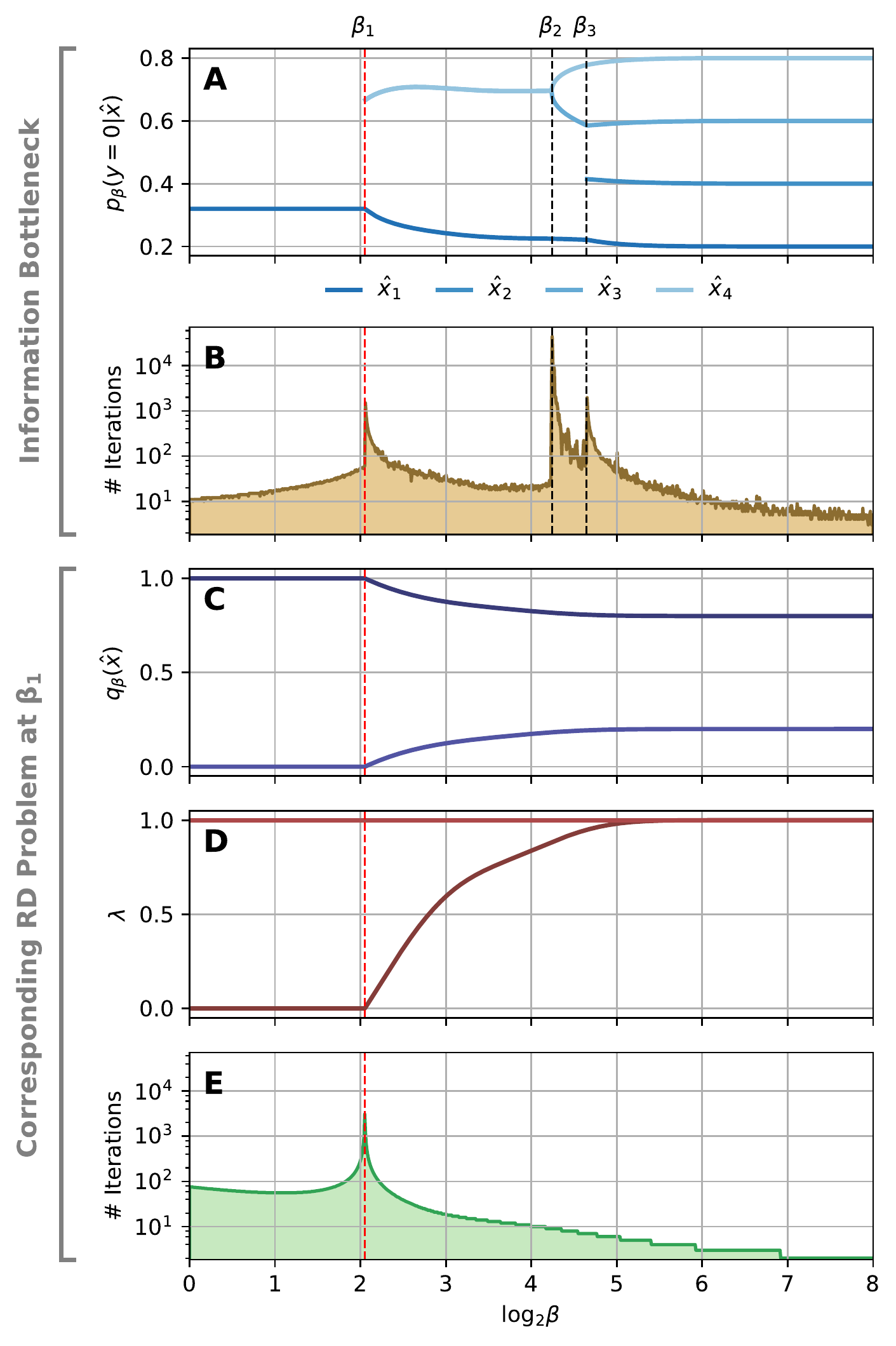}
	\caption{A simple IB problem with a binary $Y$, defined by $p(x) = (0.7,\,0.1,\,0.1,\,0.1)$ and $p(y=0 \mid x) = (0.2,\,0.4,\,0.6,\,0.8)$, as well as its corresponding RD problem. \textbf{A.} The decoder distribution, $p_\beta(y=0 \mid \hatx)$: notice the three phase transitions of the IB problem at $\beta_1,\,\beta_2,\,\beta_3$ (marked with dashed lines). \textbf{B.} Number of iterations until convergence of the IB solution ($\varepsilon = 10^{-7}$, logarithmic scale): a slowing down of at least one order of magnitude is clearly observed at each critical point. \textbf{C.} Values of $q_\beta(\hatx)$ of the corresponding (tangent) RD problem at $\beta_1$ (red dashed line): its topological transition is at $\beta_1$, as in the IB. \textbf{D.} Eigenvalues of $A$ at $q_\beta$. \textbf{E.} Number of iterations until convergence of the RD solution ($\varepsilon = 10^{-7}$, logarithmic scale): notice the single slowing down at $\beta_1$.}
	\label{fig:slowingdownIB}
\end{figure}

IB problems can be solved numerically using a modified version of the AB algorithm~\cite{IB1999}, which iterates also the decoder equation \eqref{eq:decoder}. Based on our RD analysis, we show that the IB alternating projection algorithm also exhibits critical slowing down near topological transitions (see Figure~\ref{fig:slowingdownIB}A,B).

While it can be analyzed directly, we can rely on the fact that the IB curve is an envelope of tangent RD curves~\cite{Bachrach2003}. Consider an IB solution $p_{\beta^*}(\hatx | x)$ at some value $\beta^*$ and let $\beta^- < \beta^* < \beta^+$ in some small neighborhood. Let $\mathcal \hatX^-$ be a set of representors of the effective cardinality of $p_{\beta^-}$, such that all their decoder distributions $p_{\beta^-}(y | \hatx)$ are different, and let $\mathcal \hatX^+$ be defined accordingly for $p_{\beta^+}$. Define $\mathcal \hatX^*$ to be the (formal) disjoint union $\mathcal \hatX^- \sqcup \mathcal \hatX^+$, and the (fixed) distortion function $d^*: \mathcal X \times \mathcal \hatX^* \to \mathbb R^+$ as
\begin{equation}
    d^*(x, \hatx) :=
    \begin{cases}
        D_{K\!L}[p(y | x) || p_{\beta^-}(y | \hatx)] & \hatx \in \mathcal \hatX^- \\
        D_{K\!L}[p(y | x) || p_{\beta^+}(y | \hatx)] & \hatx \in \mathcal \hatX^+
    \end{cases}
\end{equation}
Let $q_\beta$ be the solution to the RD problem defined by $\mathcal \hatX^*$ and $d^*$. Since $p_{\beta^-}$ is optimal at $\beta^-$, being the IB solution there, we know that $\supp q_{\beta^-} = \mathcal \hatX^-$; similarly $\supp q_{\beta^+} = \mathcal \hatX^+$. Notice that if the IB problem undergoes a transition at $\beta^*$, then the effective cardinality of $p_{\beta^-}$ differs from that of $p_{\beta^+}$, and therefore by Definition~\ref{def:phasetransition}, the tangent RD problem above must have a topological transition at some $\beta^-<\beta<\beta^+$. 

Taking the limits  $\beta^- \to \beta^*$ from below and $\beta^+ \to \beta^*$ from above, we say that the RD problem defined by $\mathcal \hatX^*$ and $d^*$ is the \emph{corresponding tangent RD problem to the original IB problem at $\beta^*$}.
Consequently, if the IB problem has a topological transition at $\beta_c$ then the corresponding tangent RD problem at $\beta_c$ must also have a critical transition at $\beta_c$ (see Figure~\ref{fig:slowingdownIB}C).

Finally, by its definition, the solution to the tangent RD problem at a given $\beta$ already achieves the optimal decoder distribution at that point. However, since the IB algorithm has to iterate additionally over the decoder distributions in order to converge to the IB solution at $\beta$, it is expected to perform at least the same number of iterations as the AB algorithm for the tangent RD solution there. Therefore, when $\beta$ is close enough to some critical point of the IB, the IB algorithm experiences similar slowing down there, as shown in Figure~\ref{fig:slowingdownIB}B,E.

\section{Discussion}
The AB algorithm for rate-distortion problems is known to converge uniformly to the optimal RD function in times that are $O(1/\varepsilon)$\cite{boukris1973upper}. Our results deal with the ratio between the number of iterations until $\varepsilon$-convergence and $-\log\varepsilon$, and show that this ratio increases significantly near critical points. While the $O(1/\varepsilon)$ bound is independent of $\beta$, that is, the constant does not increase near critical points, it corresponds to a much slower convergence, for sufficiently small $\varepsilon$. Moreover, our results address the convergence of the encoder $p(\hatx|x)$, not just the rate $R(D)$. At the critical points there might be different competing optimal solutions at the same $R(D)$.


For similar reasons, variational approximations to either RD or IB (e.g.~\cite{alemi2019deep}) may not suffer from CSD, as they can be too far from the optimal encoder or use different dynamics. It is not clear if other local converging algorithms, such as stochastic gradient decent, should also exhibit CSD near topological representation transitions, but we know that near the critical points the Hessian matrix of the Lagrangian at the optimum becomes singular. Thus any gradient based optimization is susceptible to CSD if it has components in the flat dimensions of the minima. 

The implications of our results to local representation learning, when effective annealing is obtained through complexity regularization, are intriguing. In such cases, deep learning in particular, the critical points can determine the location of the final representations along the optimal RD or IB curves. 

\bibliographystyle{IEEEtran}
\bibliography{isit_biblio}
\appendices

\section{Derivation of $A$} \label{sec:A}
Recall that $Z(x, \beta) := \sum_{\hatx} p_\beta(\hatx) e^{-\beta d(x, \hatx)}$, therefore
\begin{equation}
    \frac{\partial Z(x, \beta)}{\partial p_\beta(\hatx)} = e^{-\beta d(x, \hatx)} \;.
\end{equation}

For $\hatx' \neq \hatx$ we have
\begin{equation}
    \frac{\partial}{\partial p_\beta(\hatx')} \left( \frac{p_\beta(\hatx)}{Z(x, \beta)} \right)
    = - \frac{p_\beta(\hatx) e^{-\beta d(x, \hatx')}}{Z(x, \beta)^2} \;,
\end{equation}
and thus
\begin{equation}
    \frac{\partial \big[F(p_\beta, \beta)\big]_{\hatx}}{\partial p_\beta(\hatx')}
    = p_\beta(\hatx) \sum_x p(x) \frac{e^{-\beta \big( d(x, \hatx) + d(x, \hatx') \big)}}{Z(x, \beta)^2} \;. \label{eq:dFdp1}
\end{equation}

In contrast,
\begin{multline}
    \frac{\partial}{\partial p_\beta(\hatx)} \left( \frac{p_\beta(\hatx)}{Z(x, \beta)} \right)
    = \frac{Z(x, \beta) - p_\beta(\hatx) e^{-\beta d(x, \hatx)}}{Z(x, \beta)^2} \\
    = \frac{1}{Z(x, \beta)} - \frac{p_\beta(\hatx) e^{-\beta d(x, \hatx)}}{Z(x, \beta)^2} \;,
\end{multline}
and then
\begin{multline}
    \frac{\partial \big[F(p_\beta, \beta)\big]_{\hatx}}{\partial p_\beta(\hatx)}
    = 1 - \sum_x p(x) \left( \tfrac{e^{-\beta d(x, \hatx)}}{Z(x, \beta)} - \tfrac{p_\beta(\hatx) e^{-2\beta d(x, \hatx)}}{Z(x, \beta)^2} \right) \\
    = 1 - \sum_x \tfrac{p(x) e^{-\beta d(x, \hatx)}}{Z(x, \beta)} +
    p_\beta(\hatx) \sum_x p(x) \tfrac{e^{-2\beta d(x, \hatx)}}{Z(x, \beta)^2} \;. \label{eq:dFdp2}
\end{multline}

When $p_\beta(\hatx) > 0$ we have by \eqref{eq:p(hatx|x)}, applying Bayes' law,
\begin{equation}
    p_\beta(x | \hatx) = \frac{p(x)}{p_\beta(\hatx)} p(\hatx | x)
    = \frac{p(x) e^{-\beta d(x, \hatx)}}{Z(x, \beta)} \;, \label{eq:p(x|hatx)}
\end{equation}
and therefore $\sum_x \frac{p(x) e^{-\beta d(x, \hatx)}}{Z(x, \beta)} = 1$. Note, however, that the right hand side of \eqref{eq:p(x|hatx)} is well defined even when $p_\beta(\hatx) = 0$. Since we are interested only in the right derivative when $p_\beta(\hatx) = 0$ (as it is in the boundary of $\Delta \mathcal \hatX$), we can refer in that case to the limit, which also satisfies $\sum_x \frac{p(x) e^{-\beta d(x, \hatx)}}{Z(x, \beta)} = 1$. Consequently, we have from \eqref{eq:dFdp2}
\begin{equation}
    \frac{\partial \big[F(p_\beta, \beta)\big]_{\hatx}}{\partial p_\beta(\hatx)}
    = p_\beta(\hatx) \sum_x p(x) \frac{e^{-2\beta d(x, \hatx)}}{Z(x, \beta)^2} \;,
\end{equation}
which together with \eqref{eq:dFdp1} gives the result in \eqref{eq:A2}.

The formula in \eqref{eq:A} is a straightforward result of \eqref{eq:A2} and \eqref{eq:p(x|hatx)}. Although simpler, it is not defined when $p_\beta(\hatx)=0$.

\section{Proof of Theorem~\ref{thm:dimker}} \label{sec:proofThmDimKer}
\begin{IEEEproof} [Proof of Lemma~\ref{lemma:standardbasis}]
    If $p_\beta(\hatx) = 0$ then by \eqref{eq:A2} the corresponding column of $A$ is zero and thus $A \mathbf e_{\hatx} = 0$. Conversely, if $A \mathbf e_{\hat x} = 0$, then by \eqref{eq:A2} we have particularly
    \begin{equation}
        p_\beta(\hatx) \sum_x p(x) \frac{e^{-2\beta d(x, \hat x)}}{Z(x, \beta)^2} = 0 \;.
    \end{equation}
    Therefore, either $p_\beta(\hatx) = 0$ or $d(x, \hat x) = \infty$ for all $x$. But the latter implies by \eqref{eq:p(hatx|x)} that $p_\beta(\hatx | x) = 0$ for all $x$, and so anyway $p_\beta(\hatx) = 0$.
\end{IEEEproof}

\begin{IEEEproof} [Proof of Proposition~\ref{prop:allbasis}]
    We prove the proposition by induction on $r$.
    
    The case $r = 1$ is trivial. Let $r = 2$ and assume by contradiction that $\mathbf e_{\hatx_1} \notin \ker A$. This means that also $\mathbf e_{\hatx_2} \notin \ker A$, otherwise $v - v_{\hatx_2} \mathbf e_{\hatx_2} \in \ker A$, which would imply that $\mathbf e_{\hatx_1} \in \ker A$. Therefore, according to Lemma~\ref{lemma:standardbasis}, both $p_\beta(\hatx_1), p_\beta(\hatx_2) > 0$.
    
    Now, since $v \in \ker A$, we have by \eqref{eq:A2} for all $\hatx$
    \begin{align}
        0 &= \sum_{\hatx'} p_\beta(\hatx') \sum_x p(x) \frac{e^{-\beta \big( d(x, \hatx) + d(x, \hatx') \big)}}{Z(x, \beta)^2} v_{\hatx'} \\
        &= \sum_x p(x) \frac{e^{-\beta d(x, \hatx)}}{Z(x, \beta)} \Big( p_\beta(\hatx_1 | x) v_{\hatx_1} + p_\beta(\hatx_2 | x) v_{\hatx_2} \Big) \;, \label{eq:r2}
    \end{align}
    where the second equality follows from \eqref{eq:p(hatx|x)}.
    Averaging over $p_\beta(\hatx)$ gives $p_\beta(\hatx_1) v_{\hatx_1} + p_\beta(\hatx_2) v_{\hatx_2} = 0$, and thus $v_{\hatx_1} = -\frac{p_\beta(\hatx_2)}{p_\beta(\hatx_1)} v_{\hatx_2}$. Plugging this result back in \eqref{eq:r2} and dividing by $p_\beta(\hatx_2)\,v_{\hatx_2} \neq 0$ we get for all $\hatx$
    \begin{equation}
        \sum_x p(x) \frac{e^{-\beta d(x, \hatx)}}{Z(x, \beta)^2} \left( e^{-\beta d(x, \hatx_2)} - e^{-\beta d(x, \hatx_1)} \right) = 0 \;,
    \end{equation}
    where we used again \eqref{eq:p(hatx|x)}.
    Substituting $\hatx = \hatx_1, \hatx_2$ in the last equation we have
    \begin{align}
        \sum_x \tfrac{p(x)}{Z(x, \beta)^2} \left( e^{-\beta d(x, \hatx_1)} e^{-\beta d(x, \hatx_2)} - e^{-2\beta d(x, \hatx_1)} \right) &=0 \label{eq:r2a}\\
        \sum_x \tfrac{p(x)}{Z(x, \beta)^2} \left( e^{-2\beta d(x, \hatx_2)} - e^{-\beta d(x, \hatx_1)} e^{-\beta d(x, \hatx_2)} \right) &=0 \label{eq:r2b}
    \end{align}
    and subtracting \eqref{eq:r2a} from \eqref{eq:r2b} gives
    \begin{equation}
        \sum_x \frac{p(x)}{Z(x, \beta)^2} \left( e^{-\beta d(x, \hatx_1)} - e^{-\beta d(x, \hatx_2)} \right)^2 = 0 \;.
    \end{equation}
    Therefore, for all $x$ we must have $d(x, \hatx_1) = d(x, \hatx_2)$, contradicting our assumption on the non-degeneracy of $d$. Consequently, $\mathbf e_{\hatx_1} \in \ker A$, and thus $v - v_{\hatx_1} \mathbf e_{\hatx_1} \in \ker A$, implying that also $\mathbf e_{\hatx_2} \in \ker A$.
    
    Finally, let $r \geq 3$ and assume the proposition holds for all $1 \leq r' < r$. If there exists $1 \leq i \leq r$ such that $\mathbf e_{\hatx_i} \in \ker A$, then $u = v - v_{\hatx_i} \mathbf e_{\hatx_i} \in \ker A$. However, $u$ has exactly $r - 1 < r$ nonzero coordinates, namely $\hatx_j$ for $1 \leq j \leq r,\ j \neq i$, and therefore by the induction hypothesis all the corresponding $\mathbf e_{\hatx_j}$ also belong to $\ker A$. Together with $\mathbf e_{\hatx_i}$ this completes the induction step.
    
    Conversely, assume by contradiction that $\mathbf e_{\hatx_i} \notin \ker A$ for all $1 \leq i \leq r$, then by Lemma~\ref{lemma:standardbasis} we have $p_\beta(\hatx_i) > 0$ for all $\hatx_i$. This implies that $A_{\hatx \hatx_i} > 0$ for all $\hatx$ and $\hatx_i$, otherwise by \eqref{eq:A2} we would get for some $\hatx$ and $\hatx_i$ that $d(x, \hatx) + d(x, \hatx_i) = \infty$ for all $x$, contradicting our assumption on the finiteness of $d$. In particular, this means that $\sum_{i=2}^r A_{\hatx \hatx_i} > 0$.
    
    Now, since $v \in \ker A$ we have $\sum_{i=1}^r A_{\hatx \hatx_i} v_{\hatx_i} = 0$ for all $\hatx$, and thus
    \begin{align}
        0 &= A_{\hatx \hatx_1} v_{\hatx_1} + \sum_{i=2}^r A_{\hatx \hatx_i} v_{\hatx_i} \label{eq:r2c} \\
        &= \sum_{i=2}^r \frac{A_{\hatx \hatx_i} A_{\hatx \hatx_1} v_{\hatx_1}}{\sum_{j=2}^r A_{\hatx \hatx_j}} + \sum_{i=2}^r A_{\hatx, \hatx_i} v_{\hatx_i} \\
        &= \sum_{i=2}^r A_{\hatx \hatx_i} \left( \frac{A_{\hatx \hatx_1} v_{\hatx_1}}{\sum_{j=2}^r A_{\hatx \hatx_j}} + v_{\hatx_i} \right) \;. \label{eq:u}
    \end{align}
    Define the vector $u \in \mathbb R^m$ such that $u_{\hatx_i} = \frac{A_{\hatx \hatx_1} v_{\hatx_1}}{\sum_{j=2}^r A_{\hatx \hatx_j}} + v_{\hatx_i}$ for all $2 \leq i \leq r$, and all its other coordinates are 0. By \eqref{eq:u} $u \in \ker A$ and it has at most $r - 1 < r$ nonzero coordinates. If there exists $2 \leq i \leq r$ such that $u_{\hatx_i} \neq 0$ then by the induction hypothesis we would have $\mathbf e_{\hatx_i} \in \ker A$, contradicting our assumption. Therefore, for all $2 \leq i \leq r$ we must have
    \begin{equation}
        v_{\hatx_i} = - \frac{A_{\hatx \hatx_1} v_{\hatx_1}}{\sum_{j=2}^r A_{\hatx \hatx_j}} \;.
    \end{equation}
    In particular this implies that $\sgn v_{\hatx_2} = \sgn v_{\hatx_3} = -\sgn v_{\hatx_1}$. Finally, we can perform the same analysis starting at \eqref{eq:r2c} by setting aside $v_{\hatx_2}$ instead of $v_{\hatx_1}$, concluding with $\sgn v_{\hatx_1} = \sgn v_{\hatx_3} = -\sgn v_{\hatx_2}$. Together with the previous result, this means that $\sgn v_{\hatx_i} = 0$ for $i=1, 2, 3$, or equivalently that $v_{\hatx_i} = 0$, contradicting our initial assumption and completing the induction step.
\end{IEEEproof}

\section{Proof of Theorem~\ref{thm:diagonalizable}} \label{sec:diagonalizable}
\begin{IEEEproof}
    First, we deal with the case in which all $p_\beta(\hatx) > 0$. Note from \eqref{eq:A2} that $A$ can be written as the product of three matrices,
    \begin{equation}
        A = B B^\intercal C \;,
    \end{equation}
    where $B_{\hatx x} = \frac{p(x)^{\nicefrac{1}{2}}}{Z(x, \beta)} e^{-\beta d(x, \hatx)}$ and $C$ is a diagonal matrix with $p_\beta(\hatx)$ in its diagonal. Therefore, we have
    \begin{equation}
        C^{\nicefrac{1}{2}} A C^{\nicefrac{-1}{2}}
        = C^{\nicefrac{1}{2}} B B^\intercal C^{\nicefrac{1}{2}}
        = (C^{\nicefrac{1}{2}} B)(C^{\nicefrac{1}{2}} B)^\intercal \;,
    \end{equation}
    meaning that $A$ is similar to a real Gram matrix, and thus diagonalizable with non-negative eigenvalues~\cite{horn2012matrix}
    .
    
    Second, assume that $p_\beta(\hatx_i) = 0$ for $i = 1, \dots, r$ -- the first $r \geq 1$ coordinates -- and $p_\beta(\hatx) > 0$ elsewhere. Let $\hatX' = \supp p_\beta$ and denote by $p'_\beta$ and $A'$ the solutions and matrix corresponding to the RD problem restricted to $\hatX'$. Note that $Z(x, \beta)$ depends only on the support, and thus
    \begin{equation}
        A = \begin{pmatrix}
            0 & \cdots \\
            0 & A'
        \end{pmatrix} \;.
    \end{equation}
    
    Since $p'_\beta(\hatx) > 0$ for all $\hatx \in \hatX'$, there exist, by the first part of the proof, an invertible matrix $P$ and a non-negative diagonal matrix $\Lambda$ such that $P^{-1}A'P = \Lambda$. Moreover, from Theorem~\ref{thm:dimker} we have $\dim \ker A' = 0$, hence none of the values in the diagonal of $\Lambda$ (that is, the eigenvalues of $A'$) is 0.
    
    Now, we have
    \begin{multline}
        \begin{pmatrix}
            I_r & 0 \\
            0 & P^{-1}
        \end{pmatrix}
        A
        \begin{pmatrix}
            I_r & 0 \\
            0 & P
        \end{pmatrix} \\
        =
                \begin{pmatrix}
            I_r & 0 \\
            0 & P^{-1}
        \end{pmatrix}
        \begin{pmatrix}
            0 & \cdots \\
            0 & A'
        \end{pmatrix}
        \begin{pmatrix}
            I_r & 0 \\
            0 & P
        \end{pmatrix} \\
        =
        \begin{pmatrix}
            0 & \cdots \\
            0 & P^{-1} A' P
        \end{pmatrix}
        =
        \begin{pmatrix}
            0 & \cdots \\
            0 & \Lambda
        \end{pmatrix} \;, \label{eq:Asim}
    \end{multline}
    where $I_r$ is the $r \times r$ identity matrix. This means that $A$ is similar to an upper-triangular matrix, and thus all its eigenvalues appear on the diagonal of that matrix, repeated according to their respective algebraic multiplicities. Since $\Lambda$ has no zeroes in its diagonal, we conclude that the algebraic multiplicity of the eigenvalue 0 of $A$ is exactly $r$. However, according to Theorem~\ref{thm:dimker} we have $\dim \ker A = r$, or equivalently, that the geometric multiplicity of the eigenvalue 0 of $A$ is also $r$.
    
    Finally, note that the standard basis row vector $\mathbf e_{\hatx_{r + i}}^\intercal$ for $i \geq 1$ is a left eigenvector of the matrix in \eqref{eq:Asim}, associated with the eigenvalue $\Lambda_{ii}$. Therefore, also for all nonzero eigenvalues of $A$, the algebraic multiplicity must equal the geometric multiplicity. Consequently the matrix $A$ is diagonalizable with real non-negative eigenvalues.
\end{IEEEproof}

	\section{Proof of Theorem~\ref{thm:worst-case-convergence-time-for-AB}}
	\label{sec:worst-case-convergence-time-for-AB}

	\newcommand{\deltaptilde}{\delta p}
	\newcommand{\deltap}{\tilde{\delta p}}


	\begin{IEEEproof}
		Denote by $\deltaptilde_k$ the deviation vector $p_k - p_\beta$ of the $k$-th iterate from the fixed point $p_\beta$, $\deltaptilde_k(\hatx) = p_k(\hatx) - p_\beta(\hatx)$ for its $\hatx$-indexed entry. For convenience, we use the $L^1$ norm in the sequel, denoted $\|\cdot\|$. The expansion of $AB$ around $p_\beta$ 
		is~\cite{folland1990taylor}
		\begin{equation}
			AB(p_\beta + \deltaptilde_k) - p_\beta = \nabla AB\big\rvert_{p_\beta} \deltaptilde_k + O(\|\deltaptilde_k\|^2) .
		\end{equation}
		That is, to first order, a single $AB$ iteration amounts to an application of the linear operator $\nabla AB\big\rvert_{p_\beta(\hatx)} = I - A^\intercal$ to the deviation. 
		Write $B(0, r)$ for the ball of radius $r$ around the origin with respect to $L^1$. Then,
		\begin{equation}		\label{eq:taylor-expansion-in-orig-coords_}
			\deltaptilde_{k+1} \in (I - A^\intercal) \deltaptilde_k + B(0, \tilde{c} \|\deltaptilde_k\|^2) ,
		\end{equation}
		where $\tilde{c} > 0$ is a constant bounding the expansion's remainder. By Theorem~\ref{thm:diagonalizable}, $A$ is diagonalizable, and so $I - A^\intercal = P \Lambda P^{-1}$ with $\Lambda $ diagonal. Multiplying \eqref{eq:taylor-expansion-in-orig-coords_} by $P^{-1}$,
		\begin{equation}
			P^{-1}\deltaptilde_{k+1} \in P^{-1}\left(P \Lambda P^{-1}\right) \deltaptilde_k + P^{-1} B(0, \tilde{c} \|\deltaptilde_k\|^2) .
		\end{equation}
		Denote $\|\cdot\|_{op}$ for the operator norm with respect to $L^1$. By its definition, $P^{-1} B(0, r) \subset B(0, \|P^{-1}\|_{op} ~r)$. Thus, exchanging coordinates $\deltap_k := P^{-1} \deltaptilde_k$ to a basis of eigenvectors,
		\begin{equation}		\label{eq:e_k+1-in-ball-by-e_k-in-thm_}
			\deltap_{k+1} \in \Lambda \deltap_k + B(0, c \|\deltap_k\|^2) ,
		\end{equation}
		for $c := \tilde{c} \cdot \|P\|_{op}^2 \cdot \|P^{-1}\|_{op}$.
		
		Denote by $\lambda_{max} = \lambda_1 \geq \lambda_2 \geq \dots \geq \lambda_n$ the eigenvalues of $I - A^\intercal$. As noted in Section~\ref{sec:slowing-down-of-AB}, they are contained in $[0, 1)$ by our assumptions. 
		Denote the $i$-th coordinate of $\deltap_k$ with respect to this basis by $\deltap_k^{(i)}$, $i = 1, \dots, |\mathcal \hatX|$. For exposition's simplicity, suppose that $\lambda_{1}$ is a simple eigenvalue, $\lambda_1 > \lambda_2$; the proof is similar otherwise\footnote{ 
			If $\lambda_{max}$ is of multiplicity $> 1$, then take $\deltap_k^{(1)}$ to be a non-zero component along some normalized $\lambda_{max}$-eigenvector, and discard the other coordinates in the $\lambda_{max}$-eigenspace. The proof follows with minor modifications.}. 
		
		Let $0 < a < \tfrac{1}{-\log \lambda_1}$.
		An upper bound for convergence is immediate, when $\lambda_1 < 1$. Choose $\mu := \exp \left( ( \tfrac{1}{\log \lambda_1} - a )^{-1} \right)$. It satisfies $\tfrac{1}{-\log \mu} = \tfrac{1}{-\log \lambda_1} + a$, and $\lambda_{1} < \mu < 1$. Then whenever $\|\deltap_{k}\| \leq \tfrac{1}{c}(\mu - \lambda_{1})$ we have
		\begin{equation}
			\|\deltap_{k+1}\| \overset{\eqref{eq:e_k+1-in-ball-by-e_k-in-thm_}}{\leq} 
			\lambda_{1} \|\deltap_{k}\| + c \|\deltap_{k}\|^2 \leq \mu \|\deltap_{k}\| \;.
		\end{equation}
		Since $\|\deltaptilde_k\| \leq \|P\|_{op}\cdot \|\deltap_k\|$, this holds whenever
		\begin{equation}		\label{eq:delta-1-def}
			\|\deltaptilde_k\| \leq \delta_1 := \tfrac{\|P\|_{op}}{c}(\mu - \lambda_{1}) .
		\end{equation}
		Therefore, at most
		\begin{equation}
			k \leq \frac{-\log \varepsilon + \log (\|P^{-1 }\deltaptilde_{0}\| \cdot \|P\|_{op})}{-\log \mu}
		\end{equation}
		iterations are then required for $\varepsilon$-convergence of $p_k$. To capture the asymptotic convergence rate 
		we divide by $-\log \varepsilon$ and take the limit to obtain
		\begin{multline}
			\lim_{\varepsilon \to 0^+} \frac{k}{-\log \varepsilon} \leq
			\lim_{\varepsilon \to 0^+} \frac{1 + \tfrac{\log (\|P^{-1 }\deltaptilde_{0}\| \cdot \|P\|_{op})}{-\log \varepsilon}}{-\log \mu} \\
			= \frac{1}{-\log \mu} = \frac{1}{-\log \lambda_1} + a .
		\end{multline}
		
		For a lower bound, choose $\eta := \exp \left( (\tfrac{1}{\log \lambda_1} + a)^{-1} \right)$. It satisfies $\tfrac{1}{-\log \eta} = \tfrac{1}{-\log \lambda_1} - a > 0$, and thus $0 < \eta < \lambda_1$. Define,
		\begin{equation}		\label{eq:rho-def}
			\rho(\deltap) := \frac{|\deltap^{(1)}|}{\|\deltap\|}
		\end{equation}
		when $\deltap^{(1)} \neq 0$, $\rho_k := \rho(\deltap_k)$. We proceed by assuming 
		\begin{equation}		\label{eq:induction-hypothesis-for-e_k-in-thm_}
			|\deltap_k^{(1)}| \geq \rho_0 \cdot\|\deltap_k\| > 0
		\end{equation}
		for all $k \geq 0$. That is, the relative weight of the first components cannot decrease beyond its initial value at $k = 0$. This shall be justified in the sequel.
		From \eqref{eq:e_k+1-in-ball-by-e_k-in-thm_},
		\begin{multline}		\label{eq:lower-bound-on-first-coord-in-thm_}
			|\deltap_{k+1}^{(1)}| \geq \lambda_1 |\deltap_k^{(1)}| - c \| \deltap_k \|^2 \overset{\eqref{eq:induction-hypothesis-for-e_k-in-thm_}}{\geq}
			\lambda_1 |\deltap_k^{(1)}| - c \tfrac{1}{\rho_0^2} |\deltap_k^{(1)}|^2 \\ =
			|\deltap_k^{(1)}| \left[\lambda_1 - \tfrac{c}{\rho_0^2} |\deltap_k^{(1)}|\right] .
		\end{multline}
		Thus, if $|\deltap_k^{(1)}| \leq \tfrac{\rho_0^2}{c} (\lambda_1 - \eta)$ then $|\deltap_{k+1}^{(1)}| \geq \eta |\deltap_k^{(1)}|$. If the above were to hold for all $k \geq 0$, then we obtain a lower bound
		\begin{equation}		\label{eq:lower-bound-in-terms-of-e_k-1}
			|\deltap_{k}^{(1)}| \geq 
			\eta^k \; |\deltap_0^{(1)}| .
		\end{equation}
		Since $|\deltap_{k}^{(1)}| \leq \|\deltap_k\| \leq \|P^{-1}\|_{op} \cdot \|\deltaptilde_k\|$, the condition $|\deltap_k^{(1)}| \leq \tfrac{\rho_0^2}{c} (\lambda_1 - \eta)$ can be replaced by the stricter
		\begin{equation}		\label{eq:delta-2-def}
			\| \deltaptilde_k \| \leq \delta_2 := \tfrac{\rho_0^2}{c \|P^{-1}\|_{op}} (\lambda_1 - \eta) .
		\end{equation}
		Since $\|\deltap_{k}\| \geq |\deltap_{k}^{(1)}|$, and $|\deltap_0^{(1)}| = \rho_0 \|\deltap_0\|$ by the definition \eqref{eq:rho-def}, then \eqref{eq:lower-bound-in-terms-of-e_k-1} implies
		\begin{equation}
			\|P^{-1}\|_{op}\cdot \|\deltaptilde_k\| \geq
			\|\deltap_{k}\| \geq \eta^k \cdot \rho_0 \|\deltap_0\| =
			\eta^k \cdot \rho_0 \|P^{-1}\deltaptilde_0\| .
		\end{equation}
		Thus, at least
		\begin{equation}		\label{eq:lower-bound-in-thm-proof_}
			k \geq \frac{-\log \varepsilon + \log (\rho_0 \tfrac{\|P^{-1}\deltaptilde_0\|}{\|P^{-1}\|_{op}})}{-\log \eta}
		\end{equation}
		iterations are required for $\varepsilon$-convergence of $p_k$. In a manner similar to before,
		\begin{multline}
			\lim_{\varepsilon \to 0^+} \frac{k}{-\log \varepsilon} \geq 
			\lim_{\varepsilon \to 0^+} \frac{1 + \tfrac{\log ( \tfrac{\rho_0\|P^{-1}\deltaptilde_0\|}{\|P^{-1}\|_{op}})}{-\log \varepsilon}}{-\log \eta} \\
			= \frac{1}{-\log \eta} = \frac{1}{-\log \lambda_1} - a .
		\end{multline}
		
		Next, we prove assumption \eqref{eq:induction-hypothesis-for-e_k-in-thm_} by induction. That is, that the relative weight of the first component cannot decrease beyond $\rho_0$. For $k=0$ this is the definition of $\rho_0$. Assuming that \eqref{eq:induction-hypothesis-for-e_k-in-thm_} holds for $k$, we shall prove that it holds for $k + 1$. i.e., we shall prove 
		\begin{equation}		\label{eq:induction-step-to-be-proven}
			|\deltap_{k+1}^{(1)}| \geq \rho_0 \cdot \|\deltap_{k+1}\| .
		\end{equation}
		To do so, it suffices to upper-bound $\rho_0 \cdot \|\deltap_{k+1}\|$ by some $u$, to lower-bound $|\deltap_{k+1}^{(1)}|$ by some $l$, and to provide a sufficient condition for $l \geq u$ to hold.
		
		Notice that \eqref{eq:lower-bound-on-first-coord-in-thm_} provides a lower bound $l$ to the left-hand side of \eqref{eq:induction-step-to-be-proven}, by using the induction assumption \eqref{eq:induction-hypothesis-for-e_k-in-thm_}. 
		To upper-bound the right-hand side of \eqref{eq:induction-step-to-be-proven},
		\begin{multline}
			\|\deltap_{k+1}\| = |\deltap_{k+1}^{(1)}| + \sum_{i=2}^n |\deltap_{k+1}^{(i)}| \\ \overset{\eqref{eq:e_k+1-in-ball-by-e_k-in-thm_}}{\leq} 
			\lambda_1 |\deltap_{k}^{(1)}| + \lambda_2 \sum_{i=2}^n |\deltap_{k}^{(i)}| + c \|\deltap_k\|^2 \\ =
			\lambda_1 |\deltap_{k}^{(1)}| + \lambda_2 \left( \|\deltap_k\| - |\deltap_{k}^{(1)}| \right) + c \|\deltap_k\|^2 \\ =
			(\lambda_1 - \lambda_2) |\deltap_{k}^{(1)}| + \lambda_2 \|\deltap_k\| + c \|\deltap_k\|^2
		\end{multline}
		Multiplying the latter by $\rho_0$ gives an upper bound $u$ to $\rho_0 \cdot \|\deltap_{k+1}\|$.
		To prove \eqref{eq:induction-step-to-be-proven}, it remains to provide a sufficient condition for $l \geq u$ to hold,
		\begin{multline}		\label{eq:lower-upper-inequality-in-thm}
			|\deltap_k^{(1)}| \left[\lambda_1 - \tfrac{c}{\rho_0^2} |\deltap_k^{(1)}|\right] \\ \geq
			\rho_0 \left\{ (\lambda_1 - \lambda_2) |\deltap_{k}^{(1)}| + \lambda_2 \|\deltap_k\| + c \|\deltap_k\|^2 \right\} .
		\end{multline}
		By the induction assumption \eqref{eq:induction-hypothesis-for-e_k-in-thm_}, $\tfrac{1}{\rho_0} |\deltap_k^{(1)}| \geq \|\deltap_k\|$, which we use to upper-bound the right-hand side of \eqref{eq:lower-upper-inequality-in-thm}. Thus, \eqref{eq:lower-upper-inequality-in-thm} is implied by the stricter,
		\begin{equation}
			\lambda_1 - \tfrac{c}{\rho_0^2} |\deltap_k^{(1)}| \geq 
			\rho_0 \left\{ (\lambda_1 - \lambda_2) + \tfrac{\lambda_2}{\rho_0} + \tfrac{c}{\rho_0^2}|\deltap_k^{(1)}| \right\} .
		\end{equation}
		This is equivalent to,
		\begin{equation}
			|\deltap_k^{(1)}| \leq \frac{\rho_0^2 (1 - \rho_0) (\lambda_1 - \lambda_2)}{c (1 + \rho_0)} .
		\end{equation}
		In a similar manner, the latter is implied by the stricter
		\begin{equation}		\label{eq:delta-3-def}
			\|\deltaptilde_k\| \leq \delta_3 := \frac{\rho_0^2 (1 - \rho_0) (\lambda_1 - \lambda_2)}{2c \|P^{-1}\|_{op} } ,
		\end{equation}
		where we have used $1 + \rho_0 \leq 2$, and $|\deltap_{k}^{(1)}| \leq \|P^{-1}\|_{op} \cdot \|\deltaptilde_k\|$.
		That is, condition \eqref{eq:delta-3-def} is sufficient for the induction step \eqref{eq:induction-step-to-be-proven} to hold.
		
		Since our discussion focuses on the convergence of the Arimoto-Blahut algorithm, we may assume that $\|\deltaptilde_{k+1}\| \leq \|\deltaptilde_{k}\|$, for all $k \geq 0$,~\cite{Cover2006}. Therefore, it suffices to require that $\|\deltaptilde_0\| \leq \delta_i$ for $i=1, 2, 3$.
		
		Finally, consider $\delta_1$ \eqref{eq:delta-1-def}, $\delta_2$ \eqref{eq:delta-2-def} and $\delta_3$ \eqref{eq:delta-3-def} as functions of $\rho_0$, $\delta_i = \delta_i(\rho_0)$, for $i=1, 2, 3$. These are polynomials of zeroth, second and third order in $\rho_0$. They are strictly positive for $0 < \rho_0 < 1$, from their definitions. Given an initial condition $p_0$, $\delta_i(\rho(\deltap_0))$ is $\delta_i$ evaluated at the relative weight $\rho$ of the first component \eqref{eq:rho-def}, at the initial deviation $\deltap_0 := P^{-1}(p_0 - p_\beta)$. By \eqref{eq:rho-def}, $0 \leq \rho(\deltap_0) \leq 1$ for any initial condition $p_0$, and so $\delta_i(\rho)$ are defined on the unit interval $[0, 1]$. 
		
		Let $B(\delta)$ be the ball of radius $\delta$ around $p_\beta$, and 
		\begin{multline}		\label{eq:one-req-for-good-initial-condition}
			\tilde{B}_i(\delta) := \Big\{p_0 \in B(\delta): \|p_0 - p_\beta\| \leq \delta_i(\rho(\deltap_0)) \Big\},
		\end{multline}
		for $i=1,2,3$. Denote,
		\begin{equation}		\label{eq:good-initial-conditions-in-ball}
			\tilde{B}(\delta) := \tilde{B}_1(\delta) \cap \tilde{B}_2(\delta) \cap \tilde{B}_3(\delta)
		\end{equation}
		That is, $\tilde{B}(\delta)$ consists of those initial conditions $p_0$ for which the conditions (\ref{eq:delta-1-def}, \ref{eq:delta-2-def}, \ref{eq:delta-3-def})	required along the proof are met. Clearly, $\tilde{B}(\delta) \subset B(\delta)$. We will show that $\tilde{B}(\delta)$ gradually fills the entire volume of $B(\delta)$ when $\delta\to 0$:
		\begin{equation}
			\lim_{\delta \to 0} \frac{\vol \tilde{B}(\delta)}{\vol B(\delta)} = 1 ,
		\end{equation}
		where $\vol S$ stands for the volume of a set $S$. 
		It suffices to show this separately for each $\tilde{B}_i(\delta)$, $i = 1, 2, 3$.
		
		Take $\tilde{B}_2(\delta)$ for example. We show that it contains a set whose volume approaches that of $B(\delta)$, as $\delta\to 0$. Consider initial conditions in the ball $B(\delta)$ by their value of $\rho(\deltap_0)$. Formally, we rewrite $\tilde{B}_2(\delta)$ as a disjoint union
		\begin{equation}		\label{eq:B2-as-disjoint-union}
			\tilde{B}_2(\delta) = \bigcup_{0\leq \rho\leq 1} \tilde{B}_2(\delta, \rho)
		\end{equation}
		over the sets
		\begin{equation}
			\tilde{B}_2(\delta, \rho) := \Big\{p_0 \in \tilde{B}_2(\delta): \rho(\deltap_0) = \rho \Big\} .
		\end{equation}
		These can be rewritten as,
		\begin{multline}		\label{eq:B2_at_angle_reduced}
			\tilde{B}_2(\delta, \rho) = \\
			\Big\{p_0 \in B(\delta): \rho(\deltap_0) = \rho \land \|p_0 - p_\beta\| \leq \delta_2(\rho(\deltap_0)) \Big\} \\
			= \Big\{p_0 \in B(\delta): \rho(\deltap_0) = \rho \land \|p_0 - p_\beta\| \leq \delta_2(\rho) \Big\} \\
			= \Big\{p_0 \in B(\min\{\delta, \delta_2\}): \rho(\deltap_0) = \rho \Big\} 
		\end{multline}
		where the first equality is by plugging in the definition \eqref{eq:one-req-for-good-initial-condition} of $\tilde{B}_2(\delta)$.
		
		Write \eqref{eq:delta-2-def} as $\delta_2(\rho) = C\cdot \rho^2$, for $C > 0$. It has a root at 0, and is otherwise positive. Thus, there are $\delta > 0$ with $\delta \leq \delta_2(\rho)$. For these $\delta$, by \eqref{eq:B2_at_angle_reduced}
		\begin{equation}
			\tilde{B}_2(\delta, \rho) = 
			\Big\{p_0 \in B(\delta): \rho(\deltap_0) = \rho \Big\} .
		\end{equation}
		Note that $\delta\leq \delta_2(\rho)$ is equivalent to $\sqrt{\nicefrac{\delta}{C}} \leq \rho$. 
		So by \eqref{eq:B2-as-disjoint-union}, $\tilde{B}_2(\delta)$ contains the set
		\begin{equation}		\label{eq:the-shape-within-2}
			\bigcup_{\sqrt{\nicefrac{\delta}{C}}\leq \rho\leq 1} \Big\{p_0 \in B(\delta): \rho(\deltap_0) = \rho \Big\} \;.
		\end{equation}
		If a particular $\delta$ value satisfies the above inequalities, then so does any smaller $\delta > 0$ value. At the limit $\delta\to 0$, $\tilde{B}_2(\delta)$ contains a union \eqref{eq:the-shape-within-2} over all $\rho$ values, except for $\rho = 0$ which is of zero-measure.
		Since the coordinates transformation $P$ is invertible, then the latter fills almost all the volume of $B(\delta)$ as $\delta\to 0$, as required for $\tilde{B}_2(\delta)$.
		
		The argument for $\tilde{B}_1(\delta)$ and $\tilde{B}_3(\delta)$ is similar.
	\end{IEEEproof}
	
\end{document}